\documentclass[aps,prb,twocolumn,groupedaddress,showpacs]{revtex4}

\usepackage{epsfig}

\begin{document}


\title{Effects of Self-field and Low Magnetic Fields on the Normal-Superconducting Phase Transition}

\author{M.~C. Sullivan}
\thanks{Present address: Department of Physics, Ithaca College, Ithaca, NY 14850}
\author{D. R. Strachan}
\author{T. Frederiksen}
\author{R. A. Ott}
\author{C. J. Lobb}
\affiliation{Center for Superconductivity Research, Department of
Physics, University of Maryland, College Park, MD 20742}



\begin{abstract}
Researchers have studied the normal-superconducting phase
transition in the high-$T_c$ cuprates in a magnetic field (the
vortex-glass or Bose-glass transition) and in zero field.  Often,
transport measurements in ``zero field" are taken in the Earth's
ambient field or in the remnant field of a magnet.  We show that
fields as small as the Earth's field will alter the shape of the
current vs. voltage curves and will result in inaccurate values
for the critical temperature $T_c$ and the critical exponents
$\nu$ and $z$, and can even destroy the phase transition.  This
indicates that without proper screening of the magnetic field it
is impossible to determine the true zero-field critical
parameters, making correct scaling and other data analysis
impossible. We also show, theoretically and experimentally, that
the self-field generated by the current flowing in the sample has
no effect on the current vs. voltage isotherms.
\end{abstract}


\maketitle


There continues to be a great deal of interest in the
normal-superconducting phase transition of the cuprate
superconductors, due in part to the accessibility of the critical
regime\cite{chris} and also to the well-understood theories
regarding the transition.\cite{ffh}  This interest has generated a
large body of work regarding the phase transition in a magnetic
field (the vortex-glass or Bose-glass transition).\cite{see-doug}
This phase transition is generally accepted to
exist,\cite{tinkham} though reserachers have found very different
results for the critical exponents $\nu$ and $z$.\cite{see-doug}
 The existence of a vortex-glass transition has been debated by some,\cite{against} and our own recent work
has suggested a more precise criterion for determining the
critical parameters if such a phase transition does indeed
exist.\cite{doug}

Much of the knowledge from the in-field transition carries over to
the zero-field transition.  This phase transition is less often
studied, although paradoxically, the existence of this phase
transition is not in doubt and the model that governs the phase
transition is better understood. Like many other second-order
phase transitions, the normal-superconducting phase transition in
zero field is expected to obey the three-dimensional (3D) XY model
with correlation length critical exponent $\nu = 0.67$.  If the
dynamics are diffusive, then the expected dynamic critical
exponent is $z=2$.\cite{ffh,hh} However, there are widely varying
experimental results in zero field. Researchers have studied the
bulk properties of $\mathrm{YBa_{2}Cu_{3}O_{7-\delta}}$ (YBCO),
properties such as the specific heat,\cite{crit} thermal
expansivity,\cite{thermex} and transport in single
crystals,\cite{yeh} and have reported critical exponents similar
to those of the 3D-XY model, while others have found both 3D-XY
and mean field exponents ($\nu = \frac{1}{2}$) in
crystals.\cite{MFandgauss} Transport measurements in thin-film
YBCO have yielded exponents similar to those predicted by 3D-XY
theory when extrapolating from high fields to zero
field,\cite{moloni} while measurements in low fields yield
exponents larger than those expected from 3D-XY theory ($\nu
\approx 1.1$, $z \approx 8.3$).\cite{lowfields}  Measurements on
$\mathrm{Bi_{2}Sr_{2}CaCu_{2}O_{8+\delta}}$ (BSCCO), a similar
hole-doped cuprate superconductor, yield similarly conflicting
results in zero field: in crystals, there has been reported a 2D
to 3D crossover\cite{rapp1} as well as a critical regime with
multiple exponents;\cite{rapp2} this multiple critical regime has
also been observed in thin films,\cite{peligrad} while other
measurements on films claim to see 3D diffusive
dynamics\cite{osborn} and still others see a 2D
Kosterlitz-Thouless transition in this material.\cite{sefrioui}

Our work on this complex topic has called into question the method
most researchers previously used to analyze the data and has
suggested a criterion for determining the existence of a phase
transition and also for determining the critical
parameters.\cite{doug} We have also pointed out the
often-overlooked effects of current noise on non-linear current
vs. voltage ($I-V$) curves,\cite{noise} and argued that the wide
range of critical exponents in films of thickness $d \gtrsim 2000$
\AA~ is due to finite size effects limiting the size of the
fluctuations.\cite{me}

In this report we continue our re-examination of this topic and
discuss the effects of low magnetic fields on transport
measurements of the 3D zero-field normal-superconducting phase
transition.  The signature of the 3D phase transition is a change
from linear behavior ($V \sim I$) at low currents above $T_c$ to
nonlinear behavior ($V \sim I^a$) below $T_c$.  We show that, in a
manner similar to current noise,\cite{noise} magnetic fields as
low as the Earth's magnetic field can change the shape of the
$I-V$ curves and even create ohmic behavior in non-linear
isotherms.  Our results on 3D samples (similar to results  found
in 2D samples)\cite{garland} indicate that if the magnetic field
is not screened, the data analysis will yield an artificially low
value for $T_c$ and inaccurate values for $\nu$ and $z$.  Given
how forgiving the scaling analysis can be,\cite{doug} this is a
possible source of the widely varying results in zero magnetic
field.  We will also discuss another possible source of magnetic
fields: the self-field generated by the current in the sample.

We have used $I-V$ curves of YBCO to examine the
normal-superconducting phase transition in zero field.  Our films
are deposited via pulsed laser deposition onto $\mathrm{SrTiO_3}$
(100) substrates.  X-ray diffraction verified that our films are
of predominately c-axis orientation, and ac susceptibility
measurements showed transition widths $\leq 0.25$ K. $R(T)$
measurements show $T_c \approx 91.5$ K and transition widths of
about 0.7 K.  AFM and SEM images show featureless surfaces with a
roughness of $\approx 12$ nm. These films are of similar or better
quality than most YBCO films reported in the literature.

We photolithographically patterned our films into 4-probe bridges
of width 8 $\mathrm{\mu}$m and length 40 $\mathrm{\mu}$m and
etched them with a dilute solution of phosphoric acid without
noticeable degradation of $R(T)$.  Our cryostat can routinely
achieve temperature stability of better than 1 mK at 90 K. To
reduce noise, our cryostat is placed inside a screened room and
all connections to the apparatus are made using shielded tri-axial
cables.

\begin{figure}
\centerline{\epsfig{file=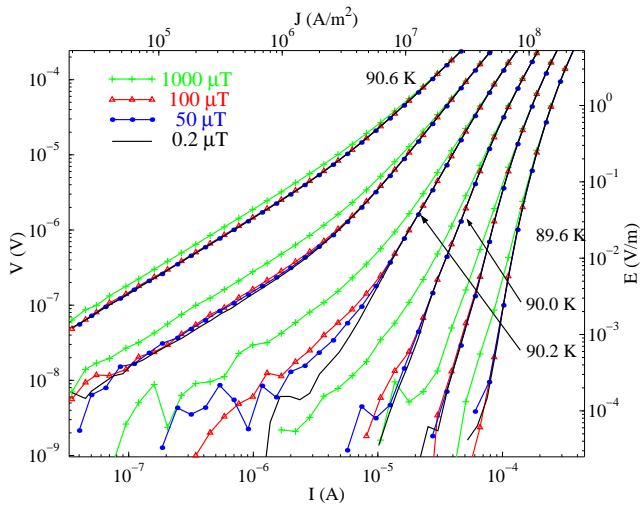,clip=,width=\linewidth}}
\caption{$I-V$ curves for a 2200~\AA~thick film with bridge
dimensions 8 $\mu$m $\times\ 40\ \mu$m.  We can see that even a
field as small as the Earth's magnetic field (50 $\mu$T) is enough
to alter the shape of the $I-V$ curves, and twice the Earth's
field (100 $\mu$T) can significantly change the shape of the
isotherms. A field of 1 mT can create ohmic tails in isotherms
which are non-linear in zero field. The $I-V$ curves are
independent of the direction of the magnetic field. The isotherms
are separated by 0.2 K.} \label{fig:selffield1}
\end{figure}

It is well-known that magnetic fields will alter the shape of the
$I-V$ curves. For this reason, we surround our cryostat with
$\mu$-metal shields to reduce the ambient field to $2 \times
10^{-7}$ T, as measured with a calibrated Hall sensor. It is
generally assumed, however, that magnetic fields on the order of
the Earth's field (50-100 $\mu$T) or even remnant fields inside
the bore of a superconducting magnet ($\sim$ 10 mT) are too small
to affect the zero-field transition, making our $\mu$-metal
shields superfluous. We have used the shields to test this
assumption, however, by attaching a copper-wire solenoid to the
outside of the vacuum can and applying small magnetic fields to
the sample.  We applied fields parallel to the c-axis and measured
the resulting $I-V$ curves.\cite{nohysteresis} These curves are
shown in Fig.\ \ref{fig:selffield1}.

In the isotherm at 90.2 K in Fig. \ref{fig:selffield1}, we can see
that a field as small as 50 $\mu$T will alter the shape of the
$I-V$ curve.  At 90.0 K and below, we can see that a field of 1 mT
(10 G) -- ten times smaller than the remnant field inside most
magnets -- can create an ohmic tail in non-linear isotherms.  We
also see in Fig.\ \ref{fig:selffield1} that magnetic fields have
the largest effect at low currents, whereas at current densities
greater than $10^8$ A/m$^2$, the magnetic field has no effect on
the isotherms. Because the clearest evidence for the transition is
expected to occur at low currents, this fact is especially
detrimental: it indicates that small magnetic fields have the
largest effect precisely where we look for the signature of the
phase transition. Thus, if the magnetic field is not screened out,
then the false ohmic tails due to small magnetic fields at low
currents will artificially \textit{decrease} $T_c$ and
consequently \textit{increase} the value for the exponent $z$
derived from the data analysis, which will in turn lead to an
inaccurate value for the exponent $\nu$.

The literature often reports measurements in ``zero field,"
however, it is unclear whether these measurements were taken
inside of a shielded cryostat or in ambient field or inside the
remnant field of a magnet.  Because of the extreme sensitivity of
the $I-V$ curves to even very small magnetic fields, this may be
an explanation for the wide range of critical exponents reported.
Moreover, our results indicate that true zero-field results may
differ even from the results taken in ambient or ``low" fields.
Finally, our results indicate that only with proper shielding of
external sources of magnetic field can you be assured of measuring
the true zero-field properties of the sample.

This result leads directly to the next question: Even if we have
shielded our cryostat of external sources, how can we be assured
of eliminating all the internal sources of magnetic field? Twisted
pair and distancing our $I-V$ bridge from the current-carrying
wires will reduce the magnetic field from the current sources,
however, there is one source of magnetic field that we cannot
eliminate: the self-field created by the current in the sample. Is
the magnetic field at the edge of the bridge large enough to
create an effect similar to that created by external fields?

\begin{figure}
\centerline{\epsfig{file=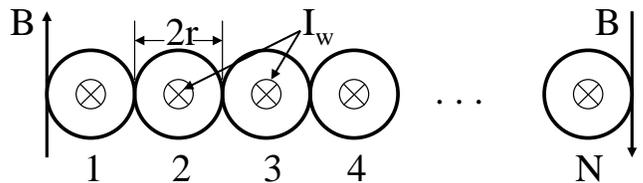,clip=,width=\linewidth}}
\caption{Approximation of the $I-V$ bridge as a collection of
$N=w/d$ wires of radius $r=d/2$.  The strongest fields occur at
the edges of the wires, denoted by \textbf{B} in the figure.}
\label{fig:selffield-schem}
\end{figure}

To answer this question, we can approximate any bridge with
thickness $d$ and width $w$ as $N=w/d$ parallel wires, each of
radius $r=d/2$ carrying current $I_w$, as sketched in Fig.\
\ref{fig:selffield-schem}.  We can then use Ampere's law to
determine the magnetic field $B$ at the edges of the bridge, as $B
= \frac{\mu_o I_w}{2 \pi r} + \frac{\mu_o I_w}{2 \pi (3r)} +
\frac{\mu_o I_w}{2 \pi (5r)}+\cdots$.  If the current density in
each wire is given by $J$, and the total current in the bridge is
$I$ (such that $I_w = I/N$), then:
\begin{equation}
B = \frac{1}{4} \mu_o d J \sum_{i=1}^{N} \frac{1}{2i-1},
\label{eq:sum}
\end{equation}
where $J = \frac{4}{\pi d w}I$.  Typical films range in thickness
from 100 nm to 300 nm and can be anywhere from 5 $\mu$m to 3 mm
wide.  Our own bridges have $w = 8$ $\mu$m and $d = 2000$ \AA,
giving $N=40$.

\begin{figure}
\centerline{\epsfig{file=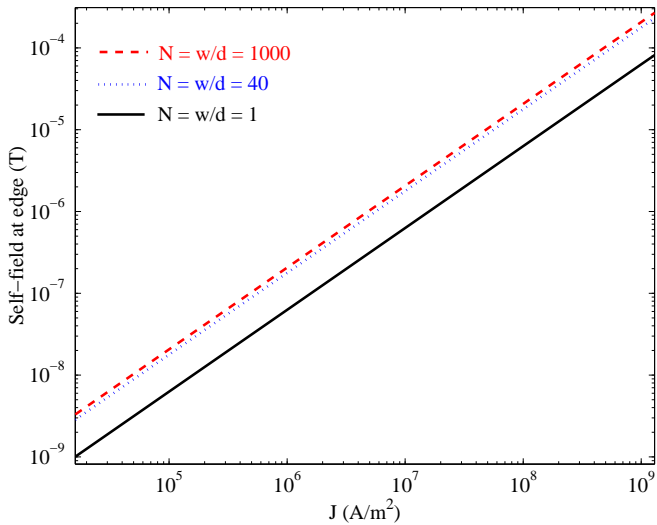,clip=,width=\linewidth}}
\caption{The self-field generated at the edges of an $I-V$ bridge
for three values of $N=w/d$.  We see that for $J \approx 10^8$
A/m$^2$, $B \approx 10$ $\mu$T, which may be large enough to
affect the sample. Our bridges have $N=40$.}
\label{fig:selffield2}
\end{figure}

We plot $B$ for three values of $N$ in Fig.\ \ref{fig:selffield2}.
We can see that at the higher current densities ($J \approx 10^8$
A/m$^2$), the self-field will create fields on the order of 10
$\mu$T. At the low current densities where the phase transition is
expected to be most apparent ($J \approx 10^6 - 10^7$ A/m$^2$),
the fields generated by the bridge are less than 1 $\mu$T.  From
this approximation, we expect the self-field to have no effect on
the $I-V$ curves of the sample.  At low currents, the self-field
is too small to have any effect, and Fig.\ \ref{fig:selffield1}
indicates even a field of 10 or 100 $\mu$T has no effect at higher
currents.

To verify that the self-field has no effect on the $I-V$ curves we
conducted an experiment based on an earlier experiment in
Josephson junction arrays\cite{jjarray} designed to reduce the
self-field at the edges of the sample.  We patterned a film of
YBCO to 8 $\mu$m $\times\ 40\ \mu$m and then covered the bridge
with photoresist and patterned a gold bridge directly above the
YBCO bridge.  The two bridges were separated by 1.1 $\mu$m of
photoresist and were not connected electrically. Using this
geometry, we can flow a given current in the YBCO bridge and the
\textit{same current} in the opposite direction in the gold
bridge. Although the field from the gold bridge will not exactly
cancel the field from the YBCO bridge due to their separation, we
can flow a higher current in the gold bridge to compensate for
this.  If the self-field generated by the YBCO bridge does create
ohmic tails at low currents, we expect the ohmic tails to
disappear when we flow a current in the opposite direction in the
gold bridge. We can also flow current in the gold bridge in the
\textbf{same} direction as the current in the YBCO bridge. In this
case, if the self-field is generating ohmic tails, we expect the
low-current ohmic tails to increase in resistance.

\begin{figure}
\centerline{\epsfig{file=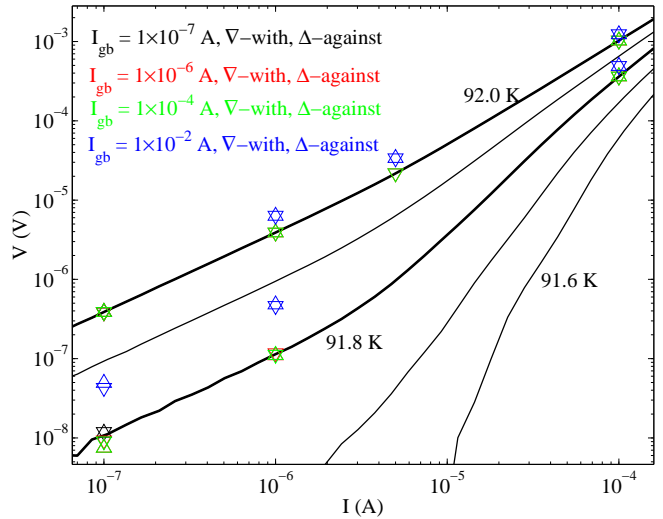,clip=,width=\linewidth}}
\caption{$I-V$ curves for a 2050~\AA~thick YBCO film with bridge
dimensions 8 $\mu$m $\times\ 40\ \mu$m.  A bridge of gold of
thickness 2000\ \AA~was patterned directly on top of YBCO film.
The gold and YBCO bridges were separated by a layer of photoresist
1.1 $\mu$m thick. The solid lines are isotherms with no current
flowing in the gold bridge. The $I-V$ points measured with
different currents in the gold bridge are represented as
triangles: $\nabla$ if the current in the gold bridge flows in the
same direction as the current in the YBCO bridge, $\triangle$ if
it flows against the current in the YBCO bridge. The colors
represent the amount of current in the gold bridge: black,
$10^{-7}$ A; red, $10^{-6}$ A; green, $10^{-4}$ A; blue, $10^{-2}$
A.  The current in the gold bridge has no effect on the $I-V$
curves until we flow 10 mA of current in the gold bridge, which
heats the YBCO bridge underneath. The isotherms are separated by
0.1 K.  The error in the points is the size of the points.}
\label{fig:selffield3}
\end{figure}

We plot the results of this experiment in Fig.\
\ref{fig:selffield3}.  In this figure, the solid lines indicate
$I-V$ curves with no current in the gold bridge.  $I-V$ points
taken with current flowing in the gold bridge are shown as
triangles.  The $\nabla$ symbol represents the gold-bridge current
flowing in the same direction as the YBCO bridge (expected to
increase the ohmic tails); the $\triangle$ symbol represents the
gold-bridge current flowing in the opposite direction as the YBCO
bridge (expected to decrease the ohmic tails).  The colors
indicate various levels of current in the gold bridge.  For
clarity, results for only two temperatures and several currents in
the YBCO bridge are presented, other temperatures and currents
yield similar results.

In Fig.\ \ref{fig:selffield3}, there is no difference between the
$\nabla$ symbols and the $\triangle$ symbols, indicating the $I-V$
points are independent of the direction of current flow in the
gold bridge.  There is also no deviation in the $I-V$ curves at
any point or at any current in the gold bridge up to 10$^{-4}$ A.
In fact, we only see an effect when we apply 10 mA to the gold
bridge, which then raises the $I-V$ points across the entire $I-V$
curve.  This is a result of the large power generated in the gold
bridge at 10 mA, the heat from which is transferred to the YBCO
bridge, raising its temperature. These results confirm the results
of Figs.\ \ref{fig:selffield1} and \ref{fig:selffield2}, namely,
that the self-field generated by the bridge is too small to
appreciably change the measured $I-V$ curve.

In conclusion, we have shown that magnetic fields as small as the
Earth's magnetic field (50 $\mu$T) can change the shape of the
zero-field $I-V$ curves and create ohmic tails at low currents.
Because a change from linear to non-linear behavior at low
currents is the expected signature of the normal-superconducting
phase transition, even small magnetic fields can lead to an
underestimate of $T_c$, an overestimate of $z$, and inaccurate
values for $\nu$.  This indicates that to measure the zero-field
phase transition correctly, the external magnetic field must be
carefully screened out. We have also examined the self-field
generated by the bridge itself, and have shown, theoretically and
experimentally, that the self-field generated by the bridge is too
small to appreciably affect the zero-field isotherms.

The authors thank D. Tobias, S. Li, H. Xu, M. Lilly, A.~J.
Berkley, Y. Dagan, H. Balci, M.~M. Qazilbash, F. C. Wellstood,
R.~L. Greene, and especially J. Higgins for their help and
discussions on this work. We acknowledge the support of the
National Science Foundation through Grant No. DMR-0302596.




\begin{thebibliography}{99}

\bibitem{chris}
C.~J. Lobb, Phys. Rev. B \textbf{36}, 3930 (1987).

\bibitem{ffh}
D.~S. Fisher, M.~P.~A. Fisher, and D.~A. Huse, Phys. Rev. B
\textbf{43},  130 (1991); D.~A. Huse, D.~S. Fisher, and M.~P.~A.
Fisher, Nature \textbf{358}, 553 (1992).


\bibitem{see-doug}
See Ref. \onlinecite{doug} for an abbreviated list of the papers
examining the normal-superconducting phase transition in a
magnetic field.

\bibitem{tinkham} M. Tinkham, \textit{Introduction to
Superconductivity}, 2nd ed. p. 356-361 (Dover, New York, 2004).

\bibitem{against}
S.~N. Coppersmith, M. Inui, and P.~B. Littlewood, Phys. Rev. Lett.
\textbf{64}, 2585  (1990); B. Brown, Phys. Rev. B \textbf{61},
3267 (2000); Z.~L. Xiao, J. H\"{a}ring, Ch. Heinzel, and P.
Ziemann, Sol. State Comm. \textbf{95},
  153  (1995);
H.~S. Bokil and A.~P. Young, Phys. Rev. Lett. \textbf{74},  3021
  (1995);
C. Wengel and A.~P. Young, Phys. Rev. B \textbf{54},  R6869
  (1996);
H. Kawamura, J. Phys. Soc. Jpn. \textbf{69},  29
  (2000);

\bibitem{doug}
D.~R. Strachan, M.~C. Sullivan, P. Fournier, S.~P. Pai, T.
Venkatesan and C.~J. Lobb, Phys. Rev. Lett. \textbf{87}, 067007
(2001); D. R. Strachan, M. C. Sullivan, and C. J. Lobb, Proc. SPIE
\textbf{4811}, 65-77 (2002).

\bibitem{hh}
P. C. Hohenberg and B. I. Halperin, Rev. Mod. Phys. \textbf{49},
435 (1977).

\bibitem{crit}
N. Overend, M. A. Howson, and I. D. Lawrie, Phys. Rev. Lett.
\textbf{72}, 3238 (1994); G. Mozurkewich, M. B. Salamon, and S. E.
Inderhees, Phys. Rev. B \textbf{46}, 11914 (1992); S. Kamal, D. A.
Bonn, N. Goldenfeld, P. J. Hirschfeld, R. Liang, and W. N. Hardy,
Phys. Rev. Lett. \textbf{73}, 1845 (1994).



\bibitem{thermex}
V. Pasler, P. Schweiss, C. Meingast, B. Obst, H. W\"{u}hl, A. I.
Rykov, and S. Tajima, Phys. Rev. Lett. \textbf{81}, 1094 (1998).

\bibitem{yeh}
N.-C. Yeh, W. Jiang, D. S. Reed, U. Kriplani and F. Holtzberg,
Phys. Rev. B \textbf{47}, 6146 (1992); N.-C. Yeh, D. S. Reed, W.
Jiang, U. Kriplani, F. Holtzberg, A. Gupta, B. D. Hunt, R. P.
Vasquez, M. C. Foote, and L. Bajuk, Phys. Rev. B \textbf{45}, 5654
(1992).

\bibitem{MFandgauss}
P. Pureur, R. Menegotto Costa, P. Rodrigues Jr., J. Schaf, and J.
V. Kunzler, Phy. Rev. B \textbf{47} 11420 (1993); A. Pomar, A.
Diaz, M. V. Ramallo, C. Torron, J. A. Veira, and Felix Vidal,
Physica C \textbf{218}, 257 (1993).

\bibitem{moloni}
K. Moloni, M. Friesen, S. Li, V. Souw, P. Metcalf, L. Hou, and M.
McElfresh, Phys. Rev. Lett. \textbf{78}, 3173 (1997).

\bibitem{lowfields}
C. Dekker, R.H. Koch, B. Oh and A. Gupta, Physica C
\textbf{185-189}, 1799 (1991); J. M. Roberts, Brandon Brown, B. A.
Hermann, and J. Tate, Phys. Rev. B \textbf{49}, 6890 (1994); T.
Nojima, T. Ishida, and Y. Kuwasawa, Czech. Jour. Phys. \textbf{46}
Suppl. S3, 1713 (1996);


\bibitem{rapp1}
S. H. Han, Yu. Eltsev, and O. Rapp, Phys. Rev. B, \textbf{57},
7510 (1998).

\bibitem{rapp2}
S. H. Han, Yu. Eltsev, and O. Rapp, Phys. Rev. B, \textbf{61},
11776 (2000).

\bibitem{peligrad}
D.-N. Peligrad, M. Mehring, and A. Dulcic, Phys. Rev. B
\textbf{69}, 144516 (2004).

\bibitem{osborn}
K. D. Osborn, D. J. Van Harlingen, V. Aji, N. Goldenfeld, S. Oh,
and J. N. Eckstein, Phys. Rev. B \textbf{68} 144516 (2003).

\bibitem{sefrioui}
Z. Sefrioui, D. Arias, C. Leon, J. Santamaria, E. M. Gonzalez, J.
L. Vicent, and P. Prieto, Phys. Rev. B \textbf{70}, 064502 (2004).


\bibitem{noise} M. C. Sullivan, T. Frederiksen, J. M. Repaci, D. R.
Strachan, R. A. Ott, and C. J. Lobb, Phys Rev B \textbf{70},
140503(R) (2004).

\bibitem{me}
M. C. Sullivan, D. R. Strachan, T. Frederiksen, R. A. Ott, M.
Lilly, and C. J. Lobb, Phys. Rev. B \textbf{69}, 214524 (2004).


\bibitem{garland}
J. C. Garland and H. J. Lee, Phys. Rev. B \textbf{36}, 3638
(1987).

\bibitem{nohysteresis} We have also reversed the magnetic field, and
found that the direction of the magnetic field (parallel or
anti-parallel to the c-axis) has no effect on the $I-V$ curve,
i.e., there is no hysteresis in the sample, as expected.

\bibitem{jjarray}
H. C. Lee, R. S. Newrock, D. B. Mast, S. E. Hebboul, J. C.
Garland, and  C. J. Lobb, Phys. Rev. B \textbf{44}, R921 (1991).


\end{thebibliography}
\end{document}